\documentclass[prb,showpacs,twocolumn,floatfix]{revtex4}   %Preprint
\usepackage{graphicx}                             %DVI/PS output
\usepackage{color}
\usepackage{amsmath}
\begin{document}
\def \be {\begin{equation}}
\def \ee {\end{equation}}
\def \nn {\nonumber}
\newcommand{\KF}[1]{\textcolor{red}{\bf [#1]}}
\newcommand{\AS}[1]{\textcolor{blue}{\bf [#1]}}
\title{Quantum Hall effect in a high-mobility two-dimensional electron gas\\on the surface of a cylinder}
\author{K. -J. Friedland$^1$}\email[Corresponding author. Electronic address: ]{kjf@pdi-berlin.de}
\author{A. Siddiki$^{2,3}$}
\author{R. Hey$^1$}
\author{H. Kostial$^1$}
\author{A.  Riedel$^1$}
\author{D. K. Maude$^4$}

\affiliation{$^1$ Paul-Drude-Institut f\"ur Festk\"orperelektronik, Hausvogteiplatz 5--7, 10117 Berlin, Germany}
\affiliation{$^2$Physics Department, Arnold
Sommerfeld Center for Theoretical Physics, and
Center for NanoScience, \\
Ludwig-Maximilans-Universit\"at, Theresienstrasse 37, 80333 Munich,
Germany} \affiliation{$^3$Physics Department,  Faculty of Arts and
Sciences, 48170-Kotekli, Mugla, Turkey} \affiliation{$^4$Grenoble
High Magnetic Field Laboratory,Centre National de la Recherche
Scientifique, 25 avenue des Martyrs, 38042 Grenoble, France}

\begin{abstract}

The quantum Hall effect is investigated in a high-mobility
two-dimensional electron gas on the surface of a cylinder. The novel
topology leads to a spatially varying filling factor along the
current path. The resulting inhomogeneous current-density
distribution gives rise to additional features in the
magneto-transport, such as resistance asymmetry and  modified
longitudinal resistances. We experimentally demonstrate that the
asymmetry relations satisfied in the integer filling factor regime
are valid also in the transition regime to non-integer filling
factors, thereby suggesting a more general form of these asymmetry
relations. A model is developed based on the screening theory of the
integer quantum Hall effect that allows the self-consistent
calculation of the local electron density and thereby the local
current density including the current along incompressible stripes. The
model, which also includes the so-called `static skin effect' to
account for the current density distribution in the compressible
regions, is capable of explaining the main experimental
observations. Due to the existence of an incompressible-compressible
transition in the bulk, the system behaves always \emph{ metal-like}  in
contrast to the conventional Landauer-B\"uttiker description, in
which the bulk remains completely insulating throughout the
quantized Hall plateau regime.
\end{abstract}
\date{\today}
\maketitle

\section {Introduction}
The self-rolling of thin pseudomorphically strained semiconductor
bilayer systems based on epitaxial heterojunctions grown by
molecular-beam epitaxy (MBE) as proposed by Prinz and
coworkers~\cite{Prinz} allows to investigate physical properties of
systems with nontrivial topology. Using a specific heterojunction,
where the high-mobility two-dimensional electron gas (2DEG) in a
13nm-wide GaAs single quantum well could be effectively protected
from charged surface states, the electron mobility in the quantum
well remains high even after fabrication of freestanding
layers~\cite{Friedland0} and particularly in semiconductor
tubes.~\cite{FriedlandI,VorobevI} Implementing this new design, the
low-temperature mean free path of electrons $l_S$ can be kept long,
comparable to the curvature radius $r$ of the tube, opening the way
to investigate curvature-related adiabatic motion of electrons on a
cylindrical surface, such as `trochoid'- or `snake'-like
trajectories.~\cite {FriedlandI,FriedlandII}
%======================================================================
\begin{figure}[!b]
\includegraphics[width=\linewidth]{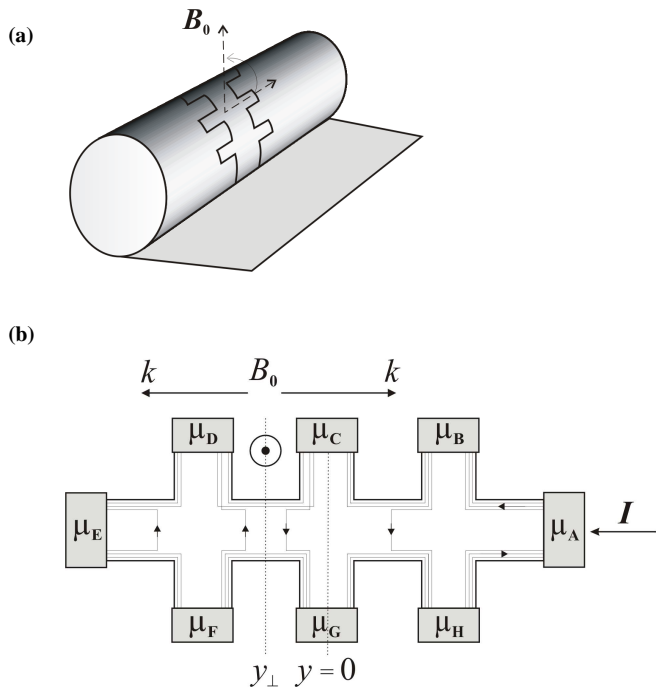}
\caption{ (a) Sketch of a  Hall-bar on the periphery of a cylinder.
(b) Schematic of such a Hall bar indicating the gradient of the
magnetic field $k$, the imposed current $I_{\rm EA}$ and imposed
chemical potentials  $\mu_i$ at leads $i$. The magnetic field is
perpendicular at the position $y_\perp$ ($y=0$ is defined to be at
the center Hall leads $C-G$). The 1DLS are shown schematically. The arrows indicate their chirality . \label{fig0} }
\end{figure}
%=========================================

Placing a tube with a high mobility 2DEG in a static and homogeneous
magnetic field $\textbf{B}_0$, the fundamental dominant modification
is the gradual change of the component of the
magnetic field  perpendicular to the surface $B_\perp$ along the periphery of the tube, which is
equivalent to a gradual change of the filling factor~$\nu$. This is
an important modification for the quantum Hall effect, which has
recently stimulated notable theoretical
interest.~\cite{Karasev,Ferrari}

Earlier investigations of the magneto-transport with spatially
varying magnetic fields, created by a density
gradient~\cite{Ponomarenko} or by magnetic field barriers inclined
with respect to the substrate facets~\cite{Ibrahim}, demonstrated
that the spatial current-density distribution is modified, thereby
creating striking lateral electric field asymmetries. Similarly, in
wave guides on cylindrical surfaces the chemical potential
differences measured along opposite edges of the Hall bar and with
opposite magnetic field directions was shown to differ by a factor
of  1000 or even to reverse its sign.~\cite{VorobevI,FriedlandII}

This large resistance anisotropy, which even persists at higher magnetic fields, was
intuitively explained by the so-called bending away of
one-dimensional Landau-states (1DLS) from the edges into the
bulk~\cite{MendachI,VorobevI}, as demonstrated in
Fig.~\ref{fig0}(b). Figure~\ref{fig0} shows schematically a Hall bar structure oriented along the periphery of a cylinder as  used for our investigation. A current $I_{\rm EA}$ is imposed between the
current leads ${E-A}$, which therefore flows parallel to the
gradient $k=\delta B_\perp/ \delta y$ and imposes the chemical
potentials $\mu_i$  at terminals  $\emph{i}$.   By adopting the Landauer-B\"{u}ttiker formalism
the longitudinal resistances can be calculated for integer filling
factors $\nu = hn (2eB)^{-1}= 1,2,3...$   as follows:

\begin{equation}\label{EquBuettigerA}
\begin{split}
R^L_{\rm DC}=~&\frac{\mu_D-\mu_C} {I_{\rm EA}}= \frac{h}{2e^2}(\frac{1}{\nu_0}-\frac{1} {\nu_{\rm DF}})= R^H_{0}-R^H_{\rm DF}\\
R^L_{\rm FG}=~&\frac{\mu_F-\mu_G} {I_{\rm EA}}= \frac{h}{2e^2}(\frac{1}{\nu_0}-\frac{1} {\nu_{\rm CG}})= R^H_{0}-R^H_{\rm CG}\\
\end{split}
\end{equation}

Here,  the position $y_\perp$ at which the magnetic field
$\textbf{B}_0$  is directed along the normal to the surface
$\emph{\textbf{n}}$, is located between the leads ${F-G}$ and
${D-C}$. $\nu_0$, $\nu_{\rm ij}$ are filling factors at the
positions $y_\perp$ and of the Hall lead pairs $i-j$, respectively.
$h$ denotes  Planck's constant and $e$  the electronic charge. For clarity, we use the superscripts $L$ and $H$ for the longitudinal and Hall resistances, respectively.
The arrows in Fig.~\ref{fig0} indicate the chirality of the 1DLS and
determine those Hall leads, from which the potential is induced into
the opposite longitudinal lead pair for a given direction of the
magnetic field. For the situation in Fig.~\ref{fig0}, the Hall
resistance $R^H_{\rm DF}$ induces a finite $R^L_{\rm DC}$, while the
Hall voltage $R^H_{\rm CG}$ do so for $R^L_{\rm FG}$, etc.

The longitudinal resistances for pairs of leads outside the position
$y_\perp$ read:
\begin{equation}\label{EquBuettigerB}
\begin{split}
R^L_{\rm CB}=~&\frac{\mu_C-\mu_B} {I_{\rm EA}}= 0 \\
R^L_{\rm GH}=~&\frac{\mu_G-\mu_H} {I_{\rm EA}}= \frac{h}{2e^2}(\frac{1}{\nu_{\rm CG}}-\frac{1} {\nu_{\rm BH}})= R^H_{\rm CG}-R^H_{\rm BH}.
\end{split}
\end{equation}
Reversing the direction of the magnetic field results in an
interchange of $R^L_{\rm DC} \rightleftharpoons R^L_{\rm FG}$ and
$R^L_{\rm CB} \rightleftharpoons R^L_{\rm GH}$ .

The resistance anisotropy in Hall bars with magnetic field gradient
along the current direction is also well known from classical
(\emph{metal-like}) electron transport studies at low magnetic fields. The anisotropy was also predicted by Chaplik, and is referred to as the
`static skin effect' (SSE). \cite{Chaplik, MendachI} An
experimental demonstration was reported by Mendach and
coworkers.~\cite{MendachII} The physical origin of this effect is
the gradual change of the Hall field along the Hall bar which acts
on the longitudinal electric field so that it becomes different on
both sides of the Hall bar. Microscopically, the SSE is a result of
an exponential current-squeezing towards one of the Hall bar edges
and is characterized by the skin length $L_{\rm skin}=
(\emph{k}\mu)^{-1} $, where $\mu$ is the carrier mobility.
Asymptotically, for high magnetic fields the SSE is described  by
the same Eq.~(\ref{EquBuettigerB}), in the  form of: $R^L_{\rm CB}=0$,
$R^L_{\rm BH}=R^H_{\rm CG}-R^H_{\rm BH}$.

Despite this similarity, both mechanisms differ antagonistically in
their microscopically origin. For the explanation of the SSE it is
assumed that a current flows exclusively at one edge of the Hall bar
which changes to the opposite one by inverting the magnetic field direction. In contrast,
the application of the Landauer-B\"{u}ttiker formalism for the 1DLS
states presupposes current flow at both edges of the Hall bar. In
the quantum Hall regime, for the situation presented in
Fig.~\ref{fig0}, the longitudinal resistance $R^L_{\rm CB}$ with
leads, which are still bound by the outermost edge channels, remains
zero at all times. In contrast, the bending of the innermost 1DLS
channels into the opposite leads causes the nonzero longitudinal
resistance $R^L_{\rm GH}$ that compensates the change of the
transverse Hall-voltages.

In this paper, we present quantum Hall effect measurements of a high-
mobility 2DEG on a cylinder surface and show that a significant part
of the results cannot be explained by the simplified 1DLS-approach.
We observe clear indications that the actual current-density
distribution in the Hall bar should be reconsidered and propose a
new model which takes into account more precisely the sequential
current flow along incompressible stripes and  \emph{metal-like} compressible
regions, for which  a current distribution according to the SSE
should be considered.

\section{Experimental}
The layer stack, with an overall thickness of 192 nm including the
high mobility 2DEG, was grown on top of a 20nm-thick $\rm In_{\rm
0.15}Ga_{\rm 0.85}As$ stressor layer, an essential component of the
strained multi-layered films (SMLF). An additional 50nm-thick AlAs
sacrificial layer is introduced below the SMLF in order to separate
the SMLF from the substrate.

For the fabrication of curved 2DEGs, we first fabricate conventional
Hall bar structures in the planar heterojunction  along the $[100]$
crystal direction. The two 20~$\mu$m-wide Hall bar arms and three
opposite 4$\mu$m-narrow lead pairs, separated by 10 $\mu$m, are
connected to Ohmic contact pads outside of the rolling area in a
similar manner as the recently developed technology to fabricate
laterally structured and rolled up 2DEGs with Ohmic
contacts.\cite{VorobevII, MendachIII} Subsequently, the SMLF
including the Hall bar was released by selective etching away of the
sacrificial AlAs layer with a $5\%$ HF acid/water solution at 4
$^\circ$C starting from a $[010]$ edge. In order to relax the
strain, the SMLF rolls up along the $[100]$ direction forming a
complete tube with a radius $r$ of about 20~$\mu$m. We report on
specific structures which are described in
Ref.~\onlinecite{FriedlandI} and which have a carrier density of
$n\cong (6.8-7.2)\times$10$^{15}$~m$^{-2}$ and a mobility of up to
90~m$^2$(Vs)$^{-1}$ along the $[100]$ crystal direction before and
after rolling-up. All presented measurements were carried out at a temperature $T$= 100~mK.

\section{Results and discussion}

\subsection {Asymmetry of the longitudinal resistances}

%======================================================================
\begin{figure}[]
%~\hspace*{-0.7cm}
\includegraphics[width=\linewidth]{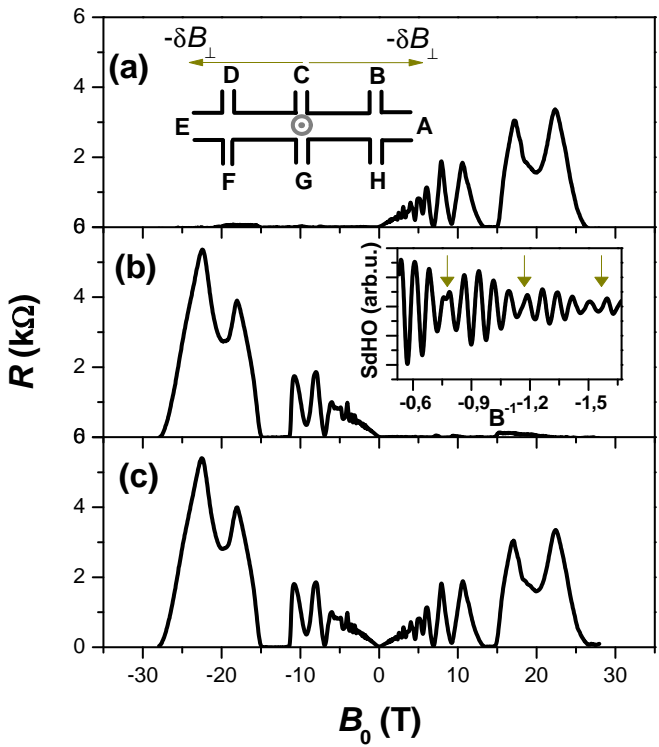}
\caption{ Longitudinal resistances with a magnetic field $B_0$
normal to the surface at the center Hall-leads $y_\perp=0$. (a)
$R^L_{\rm DC}$, the inset shows the orientation of the Hall-bar, (b)
$R^L_{\rm CB}$, the inset shows the second derivative of $R^L_{\rm CB}$,
as a function of the reciprocal of the magnetic field, (c) $R^L_{\rm
DB}$.\label{fig1} }
\end{figure}
%=========================================
The strong asymmetry of the longitudinal resistances for the current
parallel to the magnetic field gradient $k$ is demonstrated in
Fig.~\ref{fig1}.  The magnetic field is perpendicular to the surface
around the center Hall lead pair $C-G$ the position of which we
define as $y_\perp=0$. The longitudinal resistances $R^L_{\rm CB}$ -
on the right side  and $R^L_{\rm DC}$ - on the left side of this
position differ strongly for a given magnetic field and are
asymmetric with respect to the direction of the magnetic field. For
example, at a magnetic field of $B_0$ = 0.66 T, where $R^L_{\rm CB}$
shows a minimum, the ratio $R^L_{\rm DC}/ R^L_{CB}$ exceeds 300. With
the deviation $\delta y$ towards either side of  the perpendicular
field position, the component of the magnetic field decreases as
$\emph{B}_{\rm \perp}=\emph{B}_{0} $cos$(\varphi_{\delta y})$, where
$\varphi_{\delta y}$ = arcsin$(\delta y/r)$.  Accordingly, the
magnetic field gradient can be calculated as $k\cong B_0 \delta
y/r^2$. When we consider the given mobility and the field value
$B_0$ = 0.66 T, we can estimate a skin length $L_{\rm skin}\cong 670~nm$ at the positions of the next left and right pairs of the Hall leads.
As the direction of current squeezing is determined by the sign of
the field gradient, we find that for positive magnetic field values,
that the current is concentrated exponentially close to the upper Hall
bar edge between the $D-C$ leads, while the current is concentrated exponentially close to the
lower Hall bar edge between the $G-H$ leads. Inverting the magnetic
field direction results in a change of the Hall bar edges for the
current flow.

In contrast, as can be seen in Fig.~\ref{fig1}(c), the longitudinal
resistances measured between leads $D$ and $B$ is nearly symmetric,
despite the fact that $R^L_{\rm DB}$ results from current flow in
different spatial areas.

\subsection {Shubnikov de Haas oscillations}

%======================================================================
\begin{figure}[]
\includegraphics[width=\linewidth]{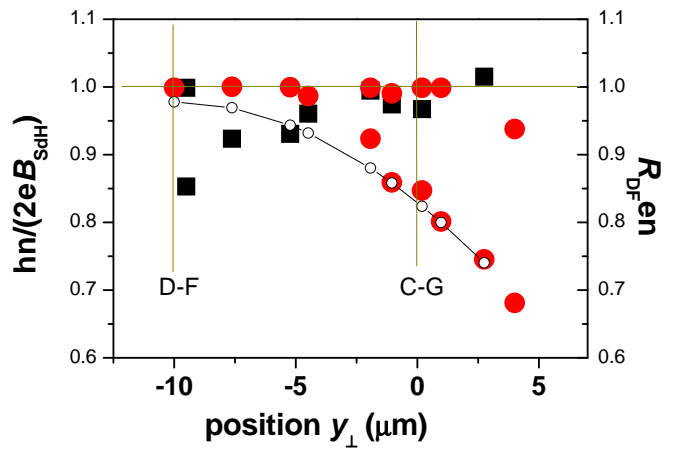}
\caption{Fundamental frequency Shubnikov de Haas oscillation $
B_{\rm SdH}^{-1}$ in units of $2e(hn)^{-1}$ calculated from $R^L_{\rm
DC}$ (closed circles) and $R^L_{\rm FG}$ (closed squares). The low
field Hall resistance $R^H_{\rm DF}$ in units of $(en)^{-1}$ is shown
by is small open circles.\label{fig2}}
\end{figure}
%======================================================================

We observe a complex structure of the Shubnikov de Haas oscillations
(SdHO). In particular, a clear beating
in the SdHO results in nodes in the second derivative of the
longitudinal resistances with respect to the inverse magnetic field
as seen, for example, in the inset of Fig.~\ref{fig1}(b). As a result,
the low-field SdHO are composed of at least  two fundamental SdHO
frequencies  $ B_{\rm SdH}^{-1}$, as calculated by a Fourier
transform analysis.

We have analyzed the two-frequency SdHO pattern by rotating the tube
around the cylinder axis through an angle $\varphi$, thereby
shifting the position $y_\perp$ away from the center pair of Hall
leads $C-G$. For $y_\perp$ values between the longitudinal voltage
leads $D-C$, Fig.~\ref{fig2} shows the dimensionless value
$hn/(2eB_{\rm SdH})$ as a function of $y_\perp$. In the same figure,
we present also the data for the classical Hall effect $R^H_{\rm
DF}en$, which corresponds nicely to the lower frequency SdHO branch.
Therefore, we conclude that this branch arises from the $B_\perp$
values at the pair of Hall leads $D-F$, which induce a voltage at
the leads $D-C$. The upper branch, close to $hn/(2eB_{\rm SdH})=1$,
reflects the SdHO for $B_0$ at the positions $y_\perp$. We conclude,
therefore, that the two-frequency SdHO pattern is in accordance with
Eq.~(\ref{EquBuettigerA}) in the form of $R^L_{\rm DC}= R^H_{0}-R^H_{\rm
DF}, R^L_{\rm FG}= R^H_{0}-R^H_{\rm CG}$ as the SdHO of the corresponding
longitudinal resistances reflects the filling factors values $\nu_0$
and $\nu_{\rm ij}$ at $y_\bot$ and the corresponding pair of Hall
leads $i-j$, respectively.

\subsection {Quantum Hall effect}

\begin{figure}[]
\includegraphics[width=\linewidth]{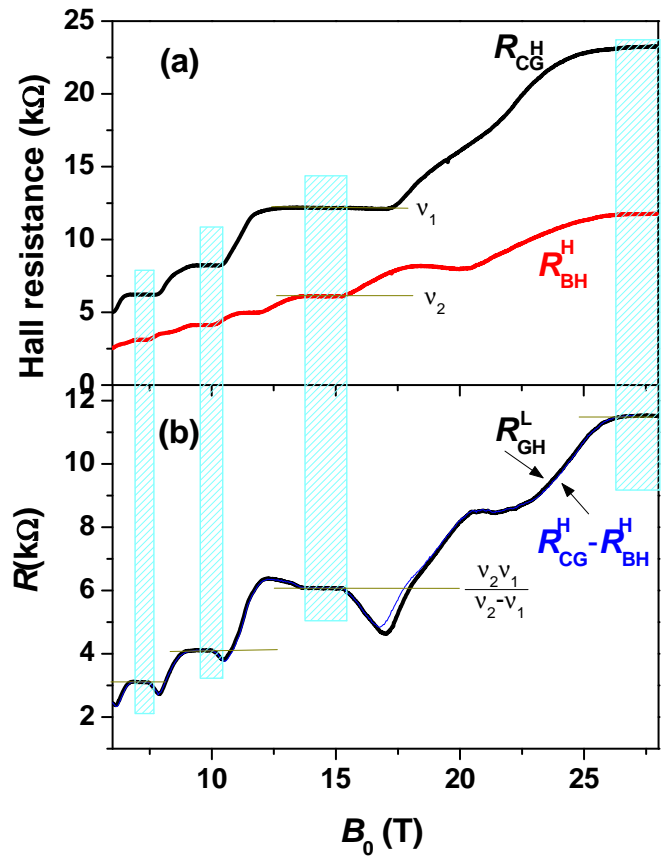}
\caption{(a) Hall resistances $R^H_{\rm CG}$ and $R^H_{\rm BH}$, (b)
longitudinal resistance $R^L_{\rm GH}$ for $y_\perp$= - 9.4 $\mu$
vs.$B_{\rm 0}$. Quantization in $R^L_{\rm GH}$ at $\nu_2 \nu_1 /(\nu_2
-\nu_1)$ is indicated by the shaded rectangles for those regions
where $R^H_{\rm CG}$ and $R^H_{\rm BH}$ are at integer filling factors.
We plot the calculated resistance $R^H_{\rm CG}$ - $R^H_{\rm BH}$ by a thin blue line.
\label{fig3}}\end{figure}

The quantum Hall effect can be observed for a wide range of magnetic
field gradients. Figure~\ref{fig3} shows the Hall resistances
$R^H_{\rm BH}$ and $R^H_{\rm CG}$ and the longitudinal resistance
$R^L_{\rm GH}$ for $y_\perp~=-9.4~\mu$m  (close to the pair of Hall
leads $D-F$), which represents a large gradient case. The filling
factors differ substantially for subsequent Hall leads. For positive
magnetic field values, the longitudinal resistances $R^L_{\rm DC}$ and
$R^L_{\rm GH}$ are always non zero. As a special case, we indicate in
Fig.~\ref{fig3} some of the magnetic field regions where both Hall
terminals are at different but integer filling factors, thus proving
the existence of quantized conductance in the non-zero longitudinal
resistance $ R_{\rm GH}$  in accordance with Eq.~(\ref{EquBuettigerB}).
Moreover, in Fig.~\ref{fig3}, it can be seen that the equation
$R^L_{\rm GH} = R^H_{\rm CG} - R^H_{\rm BH}$ holds for all positive
magnetic fields values, i.e. also for non-integer filling factors,
which is not guarantied by the Landauer-B\"{u}ttiker approach for
Eq.~(\ref{EquBuettigerB}), but is in  agreement with the local
Kirchhoff's law of voltage distribution in electronic circuits with
current.
%======================================================================
%======================================================================
%======================================================================
\begin{figure}[]
\includegraphics[width=\linewidth]{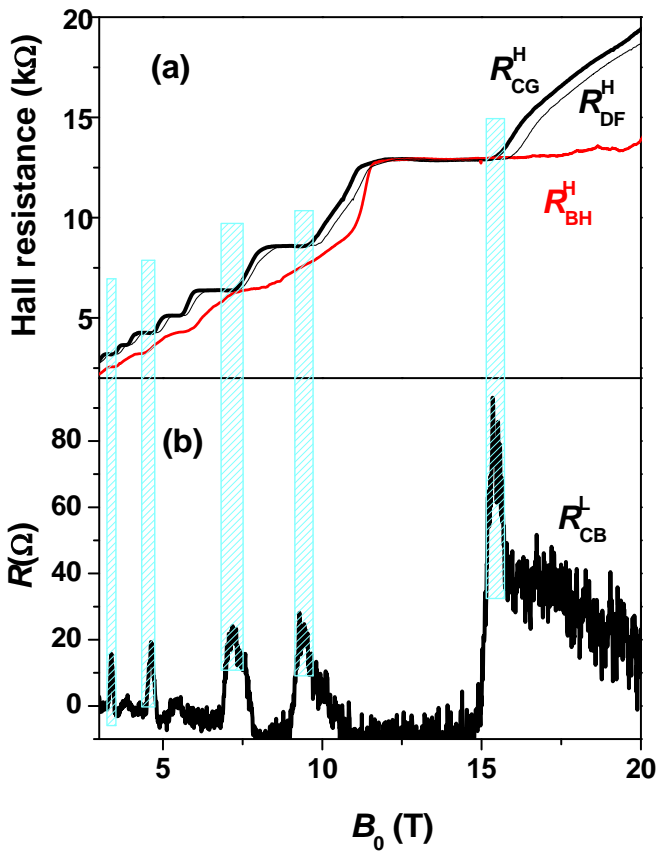}
\caption{(a) Hall resistances $R^H_{\rm CG}$, $R^H_{\rm DF}$ and $R^H_{\rm
BH}$, (b) longitudinal resistance $R^L_{\rm CB}$ for $y_\perp$=-2.1
$\mu$m vs. $B_{\rm 0}$. Peaks in $R^L_{\rm CB}$ appear at the nearest
Hall lead pair at the high magnetic field end of the quantized Hall
plateau.\label{fig4}}
\end{figure}
%======================================================================
Therefore, we conclude that for the large gradient case the
equality between the outer left and outer right expressions in
Eqs.~(\ref{EquBuettigerA}) and (\ref{EquBuettigerB}) account  for
the current and voltage distribution in our system in a more general
fashion than the simplified Landauer-B\"{u}ttiker approach for
conductance along one-dimensional channels. We will show that our
model can be used for a more quantitative explanation.

In the case of moderate gradients, i.e. small distances of $y_\perp$
from the corresponding middle pair of Hall leads, we observe a
striking deviation from the set of Eq.~(\ref{EquBuettigerB}). Despite
the fact that we should expect $R^L_{\rm CB}$=0 for any field value,
we observe  clear resistance maxima, which
even increase in height with increasing magnetic field
at the high magnetic field end of the quantized Hall plateau measured for the nearest pair of Hall leads, see
Fig.~\ref{fig4}. While the maximum values in $R^L_{\rm CB}$ remain an
order of magnitude lower then the reverse one, namely $R^L_{\rm GH}$,
they exceed the background minima due to the SSE at low magnetic
fields by an order of magnitude. We exclude that these resistance
maxima arise from a certain inaccuracy in the lead fabrication
process, which could result in a small cross talk from the voltage
inducing Hall lead pair $D-F$ into the lead C, by ensuring that the
Hall resistance $R^H_{\rm DF}$ remains quantized at corresponding
magnetic fields, see Fig.~\ref{fig4}. In order to explain this
effect, we will use our model as discussed in the following section.

\section{ Model}
We now discuss our experimental findings in the light of
self-consistent calculations of the density distribution. We exploit
the inherent similarity of the filling factor gradient generated by
the inhomogeneous magnetic field to the density gradient and utilize
current confinement to one of the Hall bar edges resulting from the
SSE. In our model calculations, we assume periodic
boundary conditions in two dimensions to describe the Hall bar
electrostatically. The magnetic field gradient is simulated by an
electron density gradient, which essentially models the filling
factor distribution over the Hall bar. The density gradient is
generated by an external potential preserving the boundary
conditions. The total electrostatic potential energy experienced by
a spinless electron is given by
\be V_{\rm tot}(x,y)=V_{\rm bg}(x,y)+V_{\rm ext}(x,y)+V_{\rm H}(x,y) \label{eq:vtot}, \ee
where $V_{\rm bg}(x,y)$ is the background potential generated by the
donors, $V_{\rm ext}(x,y)$ is the external potential resulting from
the gates (which will be used to simulate the filling factor
gradient) and the mutual electron-electron interaction is described
by the Hartree potential $V_{\rm H}(x,y)$. We assume that this total
potential varies slowly over the quantum mechanical length scale,
given by the magnetic length $l_b=\sqrt{\hbar/m \omega_c}$ so that
the electron density can be calculated within the Thomas-Fermi
approximation in 2D~\cite{SiddikiMarquardt,Siddiki08:125423}
according to
\be n_{\rm el}(x,y)=\int D(E,x,y)f(E+V_{\rm tot}(x,y)-\mu^*) dE \label{eq:tfadensity}\ee
where $D(E,x,y)$ is
the (local) density of states, $f(E)=1/[\exp(E/k_bT)+1]$ the Fermi
function, $\mu^*$ the electrochemical potential (which is constant
in equilibrium), $k_B$  Boltzmann's constant, and $T$ the
temperature. Since the Hartree potential explicitly depends on the
electron density via
\be V_{\rm H}(x,y)=\frac{2e^2}{\bar{\kappa}}\int_{A}K(x,y,x',y')n_{\rm
el}(x',y')dx'dy' ,\label{eq:tfapotential}\ee
where $\bar{\kappa}$ is an average dielectric constant ($=12.4$ for GaAs) and $K(x,y,x',y')$ is the solution of the 2D Poisson equation satisfying the
periodic boundary conditions we assume ~\cite{Morse-Feshbach53:1240} Eqs
\ref{eq:vtot} and \ref{eq:tfadensity} form a self-consistent loop,
which has to be solved numerically.

In our simulations, we start with a sufficiently high temperature to
assure convergence and decrease the temperature step by step. In the
first iteration, we assume a homogeneous background (donor)
distribution $n_0$ and calculate $V_{\rm bg}(x,y)$ from
Eq.~(\ref{eq:tfapotential}) replacing $n_{\rm el}(x',y')$  by this
constant distribution. The density gradient is produced by employing
a periodic external potential $V_{\rm ext}(x,y)=V_0\cos(2\pi
x/L_y)$, where $L_y$ is the length of the Hall bar and $V_0$ the
amplitude, reproducing also the cosine-like dependence of the
perpendicular component of the magnetic field $\emph{B}_{\rm \perp}$,  which exactly models the experimental situation represented in the
Fig.~\ref{fig1}. Here we should note that, due to the computational limitations, we
confined our calculations to a rather narrow sample. Nevertheless,
our results are scalable~\cite{SiddikiMarquardt,Siddiki08:125423} to
larger unit cells, which is, however time consuming.

\begin{figure}[]
\includegraphics[width=\linewidth]{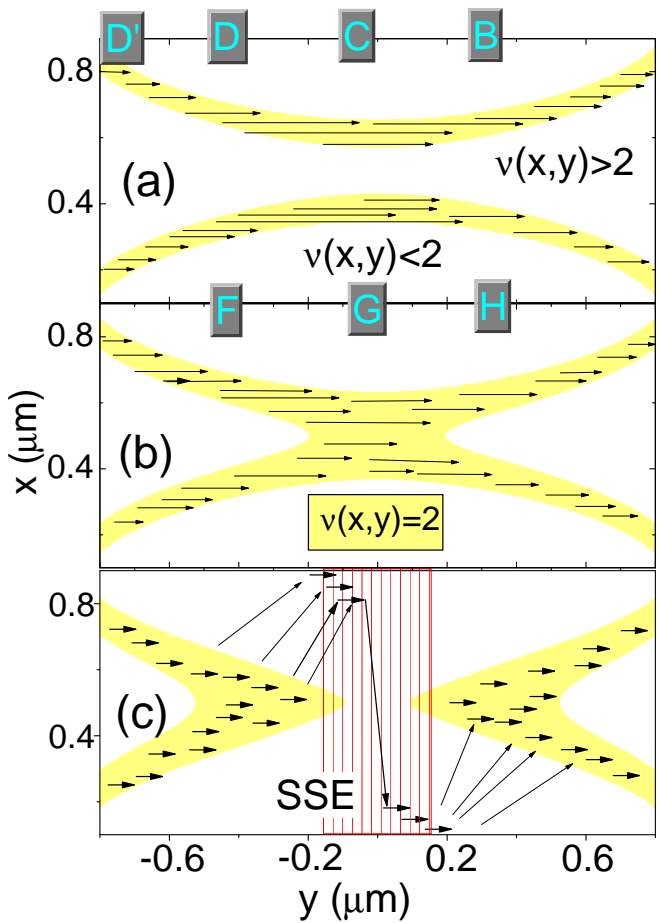}
\caption{The calculated spatial distribution of local filling
factors with integer values, $\nu(x,y)=2$ incompressible, (yellow)
and compressible, (white) for three selected values of the magnetic
field  corresponding to a filling factor at the $y_\perp=0$ position $\nu=$  (a) 2.5,
(b) 2.1,  and (c) 1.9  at 1.6 Kelvin . The unit cell is chosen to be
$1\times 2$ $\mu$m$^2$, spanning $48\times 96$ mesh points in
our numerical simulation. \label{fig5}}
\end{figure}

 As it was shown earlier for homogeneous and constricted 2DEG systems
  the calculations reveal that the wave guide is divided into compressible
bulk regions and incompressible stripes \cite{Siddiki2004}. Figure~\ref{fig5}
presents the calculated spatial distribution of the incompressible
stripes (yellow areas) for three characteristic values of the
magnetic field as a function of lateral coordinates. Arrows indicate
the current distribution, which will be discussed in detail below.
Before proceeding with the discussion of the relation between
incompressible stripes and quantized Hall effect, we would like to
emphasize the difference in the distribution of the incompressible
stripes for the selected magnetic fields.

%======================================================================
%======================================================================

In Fig.~\ref{fig5}(a), two incompressible stripes appear along the
edges of the Hall bar, which are slightly curved towards the center
due to the simulated bending, i.e. the external potential $V_{\rm
ext}(x,y)$. The two stripes merge at the center of the Hall bar at a
higher magnetic field, $\nu=2.1$, so that the center becomes
completely incompressible. Whereas, at the highest magnetic field
value considered here the center becomes compressible. In addition
to the difference between the screening properties of the
metal-like compressible (nearly perfect) and insulator-like
incompressible regions (very poor),~\cite{Siddiki03:125315} their
transport properties are also remarkable different. As mentioned
before, the compressible regions are metal-like. Therefore,
scattering is finite, and hence resistance is also finite. However,
at the incompressible stripes, the
resistance vanishes somewhat counter intuitively  since the conductance is also zero.\cite{Siddiki2004} A simple way of understanding this
phenomenon is to consider the absence of backscattering within the
incompressible stripes. Moreover, a simultaneous vanishing of both
the longitudinal resistance  and conductance is a general feature of
two-dimensional systems subjected to a strong perpendicular magnetic
field. Based on these arguments, the important features of the
integer quantized Hall effect and local probe
experiments~\cite{Ahlswede02:165} can be
explained.\cite{Siddiki04:condmat,Siddiki:ijmp}

\begin{figure}[]
\includegraphics[width=\linewidth]{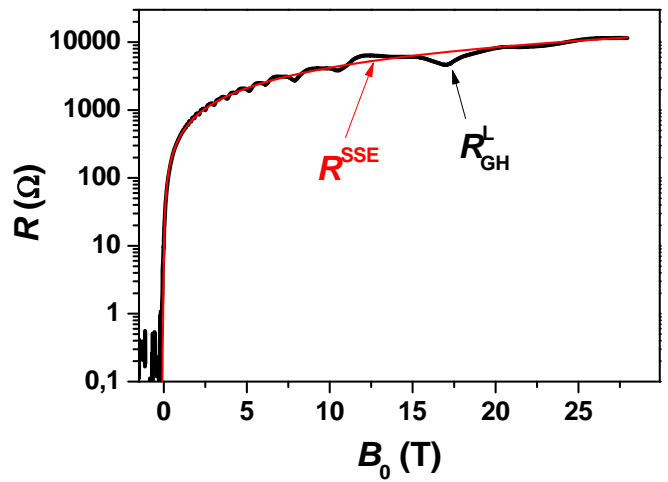}
\caption{The measured longitudinal resistance $R^L_{\rm GH}$ (black
thick line) vs. $B_{\rm 0}$ and the calculated theory curve $R^{\rm
SSE}$ (red thin line).\label{fig6}}
\end{figure}

The appearance of a metal-like compressible region along the current
path, see Fig.~\ref{fig5}(c) forces us to include another important
ingredient in our model, namely the SSE. This phenomenon is
fundamental. A fixed current  imposed in a bent \emph{metal}
stripe in a magnetic field becomes confined to one edge of
the metal due to the curvature of the system. The  following
two-parameter expression  may be derived using the SSE theory:
\be R^{\rm SSE}=R^{\rm SSE}_0\frac{B}{B_1}\frac{ e^{B/B_1}}{1-e^{B/B_1}}, \ee
where $R^{\rm SSE}_0$ is the resistance at $B_0$ =0, $B_1= r/(\mu w)$, $w$ is
the Hall bar width. In Fig.~\ref{fig6}, we provide a
semi-logarithmic plot, fitting the measured longitudinal resistance
$R^L_{\rm GH}$ of the high resistance branch with $R^{\rm SSE}$. The
fit parameters $R^{\rm SSE}_0$=6.03 and $B_{\rm 1}$=0.015 T hold for low
as well as high magnetic fields. In addition, they are very close to
the corresponding values calculated by using the given mobility, the
tube radius, and the width of the tube. We see, that the fitted curve
follows the experimental results fairly well. In particular, at low
fields, the agreement is nearly perfect since at higher filling
factors the transition from compressible to incompressible (in other
words metal to insulator) states at the center occurs over a very
narrow magnetic field range so that the bulk remains almost always
compressible. However, at higher fields the measured resistance
exhibits oscillations around the theoretical curve, which are clear
signatures of a compressible to incompressible transition in the bulk.
%======================================================================
%======================================================================

Now, we can reconsider the current distribution in our model. As
mentioned above the applied external current is confined to the
incompressible stripes, due to the absence of backscattering. In
a conventional Hall bar geometry, if an incompressible stripe
percolates from source to drain contact, the system is in the
quantized Hall regime, i.e. the longitudinal resistance vanishes and
simultaneously the Hall resistance is quantized. Such a situation is
observed in Fig.~\ref{fig5}(a), where the longitudinal resistance
measured between the leads $D-C$ (or similarly $F-G$, $C-B$, $G-H$)
vanishes, while at the same time the Hall resistance is quantized,
according $R^H_{\rm DF}=R^H_{\rm CG}=R^H_{\rm BH}= e^2/(2h)$.
Similarly, if the center becomes incompressible, Fig.~\ref{fig5}(b),
the  Hall resistance remain quantized etc. Note that now, when the
higher end of the quantized Hall plateau is approached, a striking
effect is observed. When the percolating incompressible stripe
breaks due to the bending of the structure, the bulk becomes
metal-like, and therefore the SSE comes now into play,
Fig.~\ref{fig5}(c).

First, let us discuss the Hall resistance measured between contacts
$D-F$: The quantized Hall effect remains unchanged, since the bulk is
well decoupled from the edges and the current is flowing from the
center incompressible region. Such an argument also holds for the
Hall resistance measured between the contacts $B-H$. Next, if we
measure the longitudinal resistance between say $D'-D$, we would
observe that the resistance vanishes due to the existence of the
percolating incompressible stripe between these two contacts.
However, if we measure $R^H_{\rm CG}$ simultaneously, we will see that
the quantization is smeared out since now the bulk behaves like an
ordinary metal. At this point, due to the SSE, the current is
diverted toward the edges of the Hall bar, e.g. to the upper edge on
the left side of the Hall bar and to the lower edge on the right
side for the one direction of the magnetic field and vice versa for
the opposite field direction. Therefore, the measured longitudinal
resistances  $R^L_{\rm GH}$ and $R^L_{\rm DC}$ will exhibit the SSE
with small deviations, resulting from the incompressible to
compressible transition. This scenario implies also that the
current will flow across the Hall bar at the position $y_\perp=0$
from one edge to the opposite one. We believe that this transition
around the  Hall leads  $R^H_{\rm CG}$ also accounts for the sharp
peak structure of the resistance around the transition point in
$R^L_{\rm CB}$ and $R^L_{\rm FG}$, cf. Fig.~\ref{fig4}. This effect
cannot be explained by the simple Landauer B\"uttiker approach and
indeed it would not simply occur in flat-gated  samples.

In the discussion above, we have argued that the SSE becomes
dominant when the center of the system is compressible and that
such a transition cannot be accounted for in the 1DLS picture,
where the bulk should always remain incompressible. The other
features explained by the 1DLS are equally well explained by the
screening theory, naturally, for the case of equilibrium. As an
important point, we should emphasize that the screening theory
fails to handle the non-equilibrium measurements performed by many
experimental groups (for a review see Ref. \onlinecite{Datta}),
since this theory is based on the assumption of a local
equilibrium. However, in our case the filling factor gradient is
NOT generated by the gates (i.e. creating non-equilibrium), but by
the inhomogeneous perpendicular magnetic field. Therefore, $\delta
\nu(x,y)$ is adiabatic, and the system remains in equilibrium.

\section{Conclusion}
The quantum Hall effect for a high-mobility 2DEG on a cylinder
surface show additional experimental phenomena, which indicate the
presence of a specific current-density distribution in the Hall bar.
The most prominent asymmetry relations hold not only for the
simplified case developed for the integer filling factors, but also
in a more general fashion including the transition regions between
integer filling factors. Indeed, the integer filling factor case
appears to be a relative rare case due to the gradual varying
filling factor over the current path.

We have briefly discussed the screening theory of the integer
quantum Hall effect and employed this theory to our system by
simulating the filling factor gradient. The electron density is
obtained self-consistently, while the (local) current distribution
is derived based on a phenomenological local Ohm's law. We have
explicitly shown that due to the transition from incompressible to
compressible states in the bulk, the system behaves metal-like.
Therefore, SSE is observed in our measurements,

This model allows us to explain the additional sharp peaks in
the resistance near the transition point, which appear in the
otherwise zero-resistance edge of the Hall bar and indicate a
peculiar current swing from one edge to the other. Such an effect cannot be explained by
the conventional Landauer-B\"uttiker formalism, since in this
picture the bulk remains completely insulating throughout the
quantized Hall plateau regime.

\begin{acknowledgments}
The authors gratefully acknowledge stimulating discussions with R.
R. Gerhardts  P. Kleinert and H.T.G. Grahn. We thank E. Wiebicke and M. H\"{o}ricke for technical
assistance. One of us, A.S., was financially supported by NIM Area
A. The work at GHMFL was partially supported by the European 6th
Framework Program under contract number RITA-CT-3003-505474.
\end{acknowledgments}

\end{document}